# Clinical Commissioning of the First Ultra-Compact Gantry-less Upright Proton Therapy System

Serdar Charyyev[1,*], Adam Johansson[1], Zhuoran Jiang[1], Veng Jean Heng[1], Cynthia Fu-Yu Chuang[1], Melanie R Piantino[2], Samuel Graham[3], Lawrie B Skinner[1], Nataliya Kovalchuk[1], Vivek Maradia[1], Xianjin Dai[1], Xuejun Gu[1], Bin Han[1], Lloyd Emmanuel Kamole Ghomsi[1], Susan M Hiniker[1], Yuan James Rao[1], Billy W Loo Jr[1], Lei Xing[1], Murat Surucu[1] and Yong Yang[1,*]

*1 Department of Radiation Oncology, Stanford University School of Medicine, Palo Alto, CA, USA*

*2 Mevion Medical Systems, Littleton, MA, USA*

*3 Leo Cancer Care, Middleton, WI, USA*

* Authors to whom any correspondence should be addressed.

Email: charyyev@stanford.edu and yongy66@stanford.edu

**Key words:** proton therapy, pencil beam scanning, upright radiotherapy, fixed beam, boron carbide, commissioning, adaptive aperture

## Abstract

**Background and Aims:** Upright proton therapy (PT) represents a major paradigm shift. This study reports the comprehensive clinical commissioning and benchmarking of a novel ultra-compact, gantry-free, upright PT system.

**Methods:** A treatment planning system (TPS) beam model was built and initially validated using measured integrated depth doses, spot profiles in air, and absolute dose calibration across 12 commissioning energies (49.3-227.1 MeV). Further dosimetric validation encompassed spot position accuracy, beam profiles, field penumbra, pristine and spread-out Bragg peak (SOBP) absolute doses, and disease site-specific plan verification. Mechanical and radiation isocentricity, laser alignment, and nozzle/aperture accuracy were rigorously evaluated alongside the translational and rotational accuracy of a 6-degree-of-freedom upright patient positioner (UPP). Comprehensive end-to-end tests, including an independent audit by the Imaging and Radiation Oncology Core (IROC) Houston, were performed to evaluate overall system accuracy.

**Results:** R90 range agreement between two independent depth-dose systems was within 1 mm across 49.3-227.1 MeV, with an 80%-80% peak width of 8.3-8.8 mm and a distal falloff of 4.5-4.7 mm. Absolute dose agreed with the TPS calculation within 1% for mono-energetic layers and 3% for SOBP fields. SOBP absolute-dose validation gamma passing rates (3%/2 mm) ranged from 91 to 100%; site-specific validation plans achieved ≥ 94%. Spot position accuracy, referenced to the central spot, was within 0.37 mm, and monitor unit linearity was within 1.2%. Aperture leaf position accuracy was within 1 mm photographically and within submillimeter tolerances by dosimetric picket-fence verification, with interleaf leakage of 0.6-1.2% and penumbra reduction with the adaptive aperture of up to 14.6 mm, greatest at the smallest air gap. Mechanical testing showed UPP translational accuracy within 1 mm, rotational accuracy within 0.3°, and laser and isocentricity agreement within 1 mm. End-to-end testing, including an independent IROC audit, confirmed accurate delivery (in-house film gamma 91-95%; IROC thermoluminescent dosimeter 0.96-1.09, film gamma ≥92%).

**Conclusions:** An ultra-compact, gantry-free upright PT system was successfully commissioned and validated, with overall mechanical and dosimetric accuracy suitable for clinical implementation while offering a unique opportunity to bring PT to a conventional linear-accelerator vault footprint.

## 1. Introduction

Pencil beam scanning (PBS) proton therapy (PT) delivers highly conformal dose distributions while sparing adjacent normal tissue relative to photon-based techniques[1–4]. Conventional PBS systems rely on a beam transport line and a rotating gantry, typically several meters in diameter, to deliver the beams from various angles around a supine patient. The massive size, extensive shielding requirements, and high capital cost associated with these gantry systems have historically been the primary barriers to the wider clinical adoption of PT[5–7]. Over the past decade, proton vendors have increasingly pursued compact single-room designs[8,9]. An even more compact alternative mounts a superconducting synchrocyclotron directly on the rotating gantry, eliminating the beam transport line entirely while retaining gantry rotation[10,11]. These gantry-mounted synchrocyclotron systems established a range-modulation approach based on an energy-selector (ES) plate stack together with a dynamic adaptive aperture (AA) for penumbra sharpening.

A parallel and more recent development is the shift toward gantry-free upright treatment delivery, where the beam remains fixed and the patient is positioned in a robotic chair and is rotated relative to the beam, instead of rotating the beam around the patient during delivery[5,12–19]. This approach has been explored broadly across photon and PT, including dedicated simulation and positioning studies for different body sites[20–28]. In fact, upright (seated to standing) isocentric patient rotation systems for multi-angle and arc radiotherapy have been integral to the earliest medical linacs[29,30]. Eliminating gantry rotation removes the largest and most expensive component of a proton delivery system. This fundamentally reduces the facility footprint, the shielded vault volume, and construction cost, while offering potential clinical advantages[31,32], including improved patient comfort and reduced organ motion in an upright posture[24,25,33–38]. Building on this rationale, upright PT has already entered clinical practice: fixed-beamline systems coupled with 6-degree-of-freedom (6-DOF) robotic chairs have been used clinically for skull-base chordoma and chondrosarcoma[39,40], esthesioneuroblastoma[41], and recurrent head-and-neck and brain malignancies[42], demonstrating strong positioning reproducibility[22,43]. These early clinical results support the rationale for pursuing fixed-beam, gantry-free proton delivery as a viable alternative treatment paradigm.

The Mevion S250-FIT (Mevion Medical Systems, Littleton, MA, USA) extends this trend by combining its ultra-compact superconducting synchrocyclotron platform[10,11] with an upright computed tomography (CT) scanner and a 6-DOF upright patient positioner (UPP)[44,45] (Leo Cancer Care Inc., Middleton, WI, USA) into a fully integrated, ultra-compact upright PT platform. The commissioning of another gantry-free upright proton system, built around a compact synchrotron rather than a synchrocyclotron, has been reported elsewhere[17]. The commissioning described herein of Stanford Proton Accelerator 1 (PA-1) represents the first clinical implementation of the S250-FIT ultra-compact, gantry-free, fixed-beam upright PBS architecture built on a synchrocyclotron platform. Demonstrating this reduced footprint in practice, PA-1 was installed within a converted conventional linear accelerator vault at our institution, measuring 11.5×8.5×3.5 m. Multiple S250-FIT systems are reportedly in installation or planning at other centers, reflecting the rapid emergence of this treatment paradigm.

Because the S250-FIT eliminates the gantry altogether, several elements of standard proton PBS commissioning methodology required adaptation. Gantry-angle-dependent output and beam-property characterization, standard for rotating-gantry systems[46–48], are not required because the beam delivery hardware never moves. Conversely, the UPP itself must be commissioned as the functional analog of the gantry, including translational and rotational positional accuracy, isocentricity across the full range of UPP rotation, and mechanical accuracy of the chair components (backrest, seat pan, heel and shin supports) that position the patient relative to the fixed beam.

The purpose of this work is to report the mechanical, dosimetric, and treatment planning system (TPS) commissioning of Stanford PA-1. This includes the methodological adaptations specific to this fixed-beam, upright architecture and the dosimetric consequences of its boron-carbide-based ES, providing a reference for future commissioning of similar systems. Facility shielding design and survey, as well as commissioning of the in-room upright CT imaging system, are reported separately[49,50] and are outside the scope of the present work.

## 2. Materials and Methods

### *2.A. The S250-FIT system*

The S250-FIT uses the same compact superconducting synchrocyclotron as the gantry-mounted S250i platform[10,11]: a 66-cm-diameter accelerating chamber enclosed by a superconducting magnet operating at 4.2 K and 2000 A, sustaining a maximum field strength of 10 T and accelerating protons to a fixed maximum energy of approximately 227 MeV. In the S250-FIT, the synchrocyclotron rests on a fixed, floor-mounted support rather than a rotating gantry, with the treatment nozzle mounted directly on the synchrocyclotron along a fixed beam axis (Figure 1). The single-focus scanning magnet is located directly downstream of the accelerator exit and is 50 cm in length along the beam axis, with a virtual focal length for both X- and Y-planes of 187 cm, determined as described in section 2.F. The magnet projects a maximum field size of 20×20 $cm^2$ at the isocenter plane, with a scanning speed of 10 m/s and a scanning position accuracy below 1 mm.

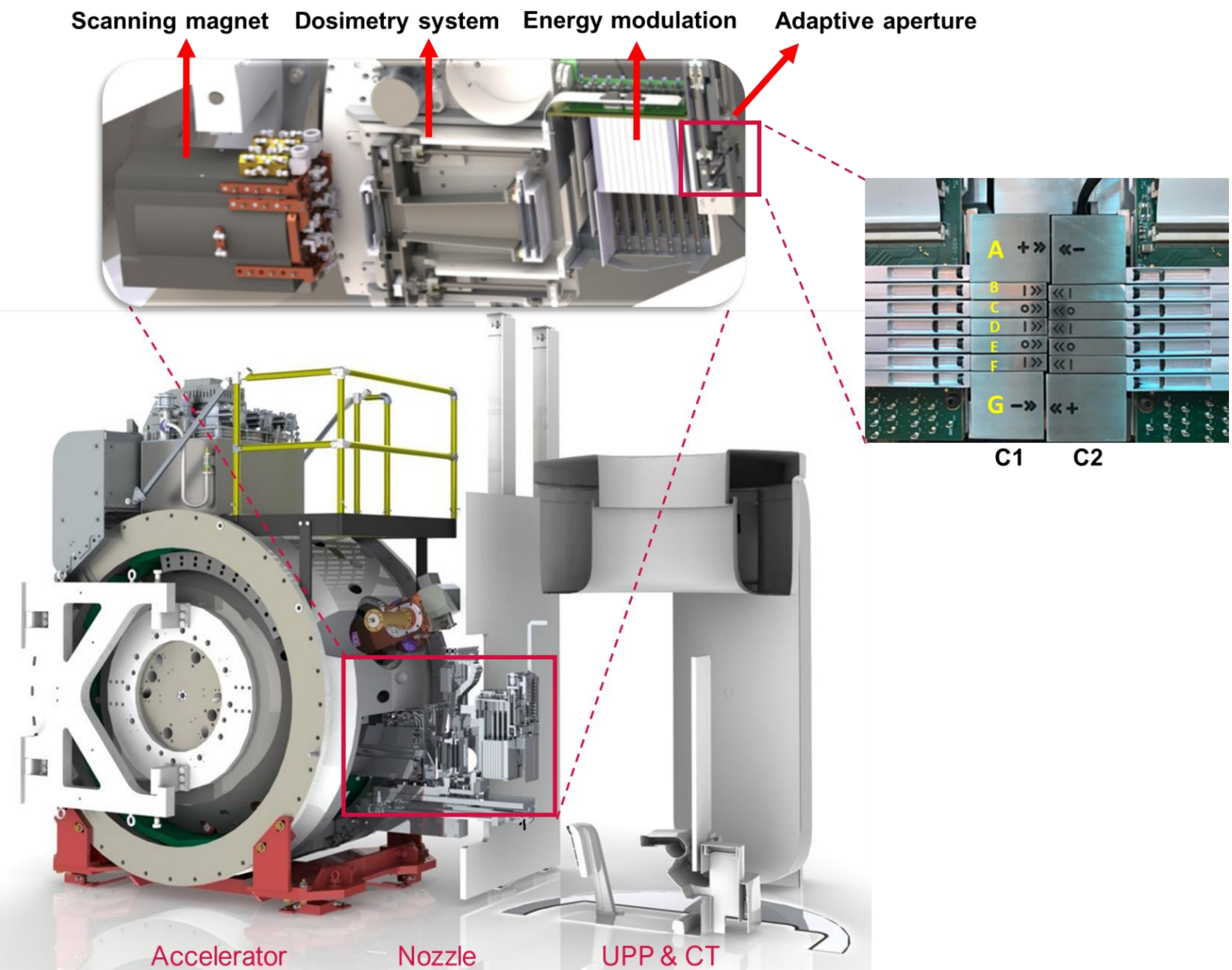

**FIGURE 1.** Overview of the S250-FIT system: the floor-mounted synchrocyclotron accelerator, nozzle assembly (scanning magnet, dosimetry system, ES, and AA) along the beam axis, the 6-DOF UPP and upright CT. AA has two banks: C1 and C2, together comprising 7 collimating pairs: A, B, C, D, E, F, G. The A and G (called jaws) are thicker (2 cm) and B, C, D, E, F (called leaves) are thinner (0.5 cm).

The treatment nozzle houses the beam monitor transmission ion chambers, the ES, and the AA, with a clinical nozzle extension range of 8.6-38.6 cm from isocenter. The ES modulates the proton energy from approximately 227 MeV down to approximately 9 MeV. The S250-FIT ES consists of 12 plates: 10 boron carbide and 2 Lexan, in contrast to the 18-plate, all-Lexan ES used in the Mevion S250i platform. Boron carbide (density ≈ 2.52 g/$cm^3$) has a higher stopping-power than Lexan (≈ 1.2 g/$cm^3$), allowing the same water-equivalent range modulation to be achieved with a shorter plate stack and a more compact nozzle. Because boron carbide and Lexan differ in atomic composition as well as density, the scattering power per unit water-equivalent thickness of the two materials is not identical; the physical thickness of degrader traversed for a given energy, and therefore the multiple-Coulomb-scattering contribution to spot size, differs from the S250i ES design. This is examined directly in the spot-size results (Section 3.B) and discussed in relation to the S250i platform in Section 4. Each ES plate is driven by an independent stepper motor, with an energy layer switching time between 50-500 ms. The AA is a dynamic beam-trimming, mini-multileaf collimator consisting of seven pairs of nickel leaves and jaws, all 10 cm thick in the beamline direction, mounted on a translating carriage, Figure 1, to position anywhere within the treatment field. The five inner leaf pairs (0.5 cm wide) incorporate a two-step tongue-and-groove design, while the two outer jaw pairs (2 cm wide) incorporate a separate interlocking bevel, both intended to reduce interleaf leakage. This design is similar to the AA previously reported[10] for the S250i platform.

The distinguishing component of the S250-FIT is the 6-DOF UPP, which replaces the rotating gantry as the means of achieving multiple treatment beam angles relative to the patient. The UPP provides three translational DOF (lateral, longitudinal, vertical) and three rotational DOF (pitch, roll, and yaw/rotation through the full 360°), together with adjustable backrest angle (±15°, in 5° increments), seat height, seat pan angle, heel stop, and shin rest configurations. Because the beam delivery hardware is stationary, beam output is intrinsically independent of treatment angle, in contrast to gantry-based systems where angle-dependent output constancy must be separately verified. Patient immobilization is achieved with upright treatment masks, belts, or vacuum bags appropriate to the treated site. TPS used is RayStation v2024A (RaySearch Laboratories, Stockholm, Sweden) with a Monte Carlo (MC) dose calculation algorithm, and the oncology information system is Aria v18 (Varian Medical Systems, Palo Alto, CA) for patient information management. An in-room upright CT scanner is used for treatment simulation and volumetric imaging at the imaging position. The treatment isocenter is located 35 cm inferior to the CT imaging isocenter and 90 cm above the floor. Currently, upright CT image cannot be acquired once positional shifts have been applied at the treatment position; however, a supplementary oblique kV-kV planar imaging system is used at the treatment position to verify that the patient has not moved after applying shifts using the CT of the day.

Table 1 summarizes the seven domains of commissioning activities followed in this work, along with their constituent sub-items.

**TABLE 1.** Domains of commissioning activities performed for the S250-FIT, with constituent sub-items for each.

|  |  |
|---|---|
| **Safety systems** | Warning lights; emergency stop; motion enable/stop; door interlocks; last man out switches; touch sensors; patient audio-visual monitoring; radiation monitors; dose reporting; secondary monitor ion chamber |
| **Non-primary radiation** | Measurements with Fluke, Swendi and neutron bubble detectors |
| **Adaptive aperture** | Position accuracy, leakage, penumbra reduction |
| **Mechanical tests** | UPP platform chair and nozzle positional accuracy; laser alignment; isocentricity |
| **Beam commissioning & validation** | IDD curves; spot profiles; absolute dose; X and Y focal lengths; off axis spot profiles and spot positioning accuracy; MU linearity; field size/penumbra/flatness/symmetry; SOBP absolute dose; SOBP depth dose profile; SOBP field size and penumbra; site-specific validation plans, end-to-end tests, IROC output irradiation and head-and-neck phantom end-to-end test |
| **2D imaging at the treatment position** | Imaging dose, image quality, isocenter accuracy |
| **Others** | Water-equivalent thickness determination |

### *2.B. Safety systems*

Facility shielding was designed and surveyed to satisfy California Department of Public Health Radiologic Health Branch requirements, 10 CFR Part 20, and NCRP guidance. The full shielding design, MC and analytic dose-equivalent calculations, and survey measurements are reported in a companion publication[50] and are not reproduced here.

Safety-interlock testing verified warning lights, emergency-stop functionality, motion-enable controls, door interlocks, last man out switches, touch sensors, patient audio-visual monitoring, radiation monitors, dose reporting, and agreement between primary and secondary beam monitor ion chamber counts.

### *2.C. Non-primary radiation*

To assess non-primary radiation, a 10×10×6 $cm^3$ spread-out Bragg peak (SOBP) field delivered 4 Gy onto a solid water phantom. At 100 cm laterally on either side of the phantom, three detectors: a neutron bubble detector, a Fluke survey meter, and a Swendi neutron/gamma detector were placed sequentially, with separate measurements performed for each. Each detector reading was converted to dose equivalent (mSv), divided by the delivered dose (4 Gy) to obtain mSv/Gy, then expressed as a percentage of the delivered dose[51,52].

### *2.D. Adaptive aperture*

AA leaf position accuracy was verified using a Mevion-supplied transparency template attached to the nozzle (exit window removed), photographed at six positions and compared against the nominal target position, with a ±1 mm tolerance. Dosimetric AA leaf positioning was additionally evaluated using an automated picket-fence-style method: a plan of 7 rectangular strip fields (in both X and Y orientation) was generated in RayStation and delivered to an IBA Lynx scintillator detector (IBA Dosimetry, Schwarzenbruck, Germany). A MATLAB-based tool was used to compare measured leaf positions against TPS calculations, as described previously[53]. AA interleaf leakage was assessed using two scans acquired with a Phoenix detector (IBA Dosimetry, Schwarzenbruck, Germany) aligned at isocenter. The first scan measured the maximum absorbed dose of an uncollimated, full-energy field with adequate water-equivalent buildup to measure at the Bragg peak. The second scan was acquired with the AA closed to block all spots from the first field, with the buildup removed so that leakage could be assessed independently of energy. The beam (as shown in Figure 2C) was directed separately through the AA jaws (position A and G, pooled across both banks) and leaves (positions B–F, pooled by bank C1 and C2), designed with an interlocking bevel and tongue-and-groove interlock, respectively, to prevent leakage. The two scans were analyzed, and leakage was determined by comparing the maximum dose at the four spot positions between scans. Moreover, AA penumbra reduction was evaluated with the PTW Octavius 1500XDR two-dimensional (2D) ion chamber array (PTW Dosimetry, Freiburg, Germany) across three energies (49.3, 154.4, 227.1 MeV), two field sizes (10×10, 20×20 $cm^2$), and two air gaps (8.6, 38.6 cm), comparing 20%-80% penumbra with and without the AA.

### *2.E. Mechanical commissioning*

Because the S250-FIT has no rotating gantry, the UPP was commissioned as the functional equivalent of gantry-angle and couch-positional testing, organized into three categories: (a) 6-DOF platform motion (lateral, longitudinal, and vertical translation; pitch, roll, and rotation/yaw), (b) chair-component motion (seat height, seat pan angle, heel stop, shin rest, and backrest angle), and (c) nozzle extension. Motion of the upright CT gantry itself (CT tilt angle and CT height) is commissioned and reported as part of the companion upright CT publication[49]. Positional accuracy of individually adjustable chair and nozzle components was verified against the control-system indicator using independent measurement devices (digital level, ruler, protractor) across the full clinical range of each component: backrest angle (-15° to +15° in 5° increments) and nozzle extension (8.6-38.6 cm) at multiple points across their range, and seat height, seat pan angle, heel stop position, and shin rest position, each at their minimum, median, and maximum settings. The 6-DOF UPP was separately verified for translational accuracy (lateral, longitudinal, vertical; full clinical range in 5-cm increments) and rotational accuracy (pitch and roll, -3° to +3° in 1° increments; rotation/yaw, 0°-330° in 30° increments). The seven-laser alignment system (central-axis, 4 transverse, 2 sagittal) was verified against the imaging isocenter and against known UPP translational offsets.

Isocentricity, the coincidence of the radiation, imaging, and mechanical isocenters, was evaluated using a Logos XRV scintillation detector (Logos Systems International, Scotts Valley, CA, USA). The detector position relative to the imaging isocenter was verified first by upright CT imaging and then by kV-kV imaging, and a proton beam was delivered to the detector with the UPP rotated in 30° increments through the full 360° range to confirm coincidence of the radiation and mechanical isocenters across UPP rotation.

### *2.F. Beam commissioning and validation*

Integrated depth dose (IDD) curves were measured for 12 commissioning energies spanning 49.3 MeV (nominal range 2.2 cm) to 227.1 MeV (nominal range 32.2 cm). 49.3 MeV was adopted as the practical lower limit for IDD acquisition due to the side-wall buildup in the water phantom at lower energies, although the ES is capable of modulating further. Two systems were used: (1) an IBA 1D water tank with a large-diameter (120 mm) parallel-plate Bragg peak chamber (IBA Dosimetry, Schwarzenbruck, Germany), scanned through single-spot beams, from which proximal R80, distal R90/R80/R20, 80%-80% peak width, and 20%-80% distal falloff were determined; and (2) an IBA Zebra multi-layer ionization chamber (MLIC) (IBA Dosimetry, Schwarzenbruck, Germany), used to establish a baseline reference for periodic constancy checks, from which range metrics were extracted using a Bortfeld analytical fit[54].

Single-spot profiles in air were measured (with 0.5 mm lateral resolution) with IBA Lynx at five positions along the central-axis (20 and 10 cm upstream of isocenter, isocenter, 10 and 20 cm downstream) for all 12 commissioning energies, with spot sigma determined from a 2D Gaussian fit in X and Y.

The X- and Y-direction focal lengths were determined from single-spot lateral separations measured with the IBA Lynx at five planes spanning -20 cm to +20 cm from isocenter, and the resulting spot separation was fit linearly against distance from isocenter for each direction independently. The focal length is defined as the distance from isocenter to the point at which the fit extrapolates to zero separation. As a cross-check, the fit was repeated using only the two extreme planes (-20 and +20 cm).

Absolute dose calibration followed the IAEA TRS-398 formalism[55] using an ADCL-calibrated IBA PPC05 parallel-plate ionization chamber and a Dose-X electrometer (IBA Dosimetry, Schwarzenbruck, Germany). For each of the 12 commissioning energies, uniform single-energy-layer plans (2.5-mm spot spacing) were delivered for four field sizes (5×5, 10×10, 15×15, and 20×20 $cm^2$) with the chamber positioned in a PTW 3D water tank (PTW Dosimetry, Freiburg, Germany) at a depth appropriate to each energy (1.0 cm for 49.3 through 114.5 MeV, 5 cm for all higher energies), and the measured dose was compared with the TPS-calculated dose for the same geometry.

Although IDD curves, spot profiles, and absolute dose were measured at 12 energies and multiple field sizes as described above, only the highest-energy (227.1 MeV) IDD, the spot profile as a function of position along the beam axis, and the 10×10 $cm^2$ field absolute dose are used to model the beam in RayStation[56]. The TPS transports the modeled 227.1 MeV beam through the ES plates to derive the full range of clinical energies rather than requiring an independent beam model at each energy. All other measurements serve to validate the resulting model and to establish baselines for periodic constancy checks.

As an additional cross-check, IDD curves and output were also measured at all energies with a PTW 3D water tank, and spot profiles were independently verified with a Phoenix detector, although none of these measurements were used to model the beam.

RayStation v2024A employs an MC dose calculation algorithm for PBS PT, commissioned using the beam model input data as described above. Validation proceeded through increasing levels of clinical complexity. First, SOBP plans were created for target volumes of 125 (5×5×5), 500 (10×10×5), 1000 (10×10×10), and 1125 (15×15×5) $cm^3$ with modulation centered at depths of 5, 10, and 20 cm (as applicable). Each SOBP was delivered without collimation and with both static and dynamic AA configurations. Dose distributions at the prescribed depth were measured with the Octavius 1500XDR array in solid water slabs (PTW Dosimetry, Freiburg, Germany) and compared with TPS calculations with gamma criteria of 3%/2 mm and a low-dose threshold of 10%. Absolute point doses at the SOBP centers were measured with PPC05 parallel-plate chamber, and the results were compared with the TPS calculation with ±3% criterion. Field size (50% isodose) and lateral penumbra (20%-80% falloff in the X and Y directions) were measured for the same SOBP geometry using the Octavius 1500XDR array.

Disease site-specific validation plans were generated for seven anatomical sites: brain (PA, RPO, LPO), C-shaped target (AP, RPO, LPO), head-and-neck (AP, RPO, LPO), pelvis (RLAT, LLAT), breast (AP), prostate (RLAT, LLAT), and craniospinal irradiation (PA superior, PA inferior). Individual beams were delivered to solid water and measured with the Octavius 1500XDR array at different depths per beam. Results were compared to TPS with point-dose and gamma analysis criteria as above.

End-to-end tests exercised the complete clinical workflow and evaluated overall target localization accuracy and dose delivery accuracy of the system: upright CT simulation, image transfer to RayStation, contouring, plan optimization and dose calculation, plan export to Aria v18, patient setup and immobilization on the UPP, image-guided alignment using upright CT and kV-kV verification at treatment position, delivery, and record-and-verify. Tests were performed on an in-house phantom capable of housing both an ion chamber and radiochromic film, with point dose verified by PinPoint ion chamber (PTW Dosimetry, Freiburg, Germany) and 2D dose verified with GafChromic (Ashland Inc., Bridgewater, NJ, USA) EBT film.

An independent external audit was performed through the Imaging and Radiation Oncology Core (IROC) quality assurance (QA) program, comprising (i) thermoluminescent dosimeter (TLD) verification of absolute machine output and (ii) an anthropomorphic head-and-neck phantom end-to-end irradiation with TLD point-dose and GafChromic film planar-dose analysis[57].

Finally, spot position accuracy was evaluated with a 25-spot test pattern (50-mm spacing) delivered to the IBA Lynx. Monitor unit (MU) linearity was assessed at 227 MeV by delivering nine MU settings (0.271-20 MU/spot) to the PPC05 chamber at 5-cm depth in solid water (tolerance ±5% for MU ≤ 5, ±2% for MU > 5). The maximum achievable field size (20 × 20 $cm^2$) was verified using a uniform field plan delivered to the Octavius 1500XDR array. Uncollimated monolayer field size, penumbra (20%-80%), flatness (Eq. 1), and symmetry (Eq. 2) were evaluated for three representative energies (49.3, 154.4, and 227.1 MeV) at the smallest (8.6 cm) and largest (38.6 cm) air gaps.

$$flatness = 100 * (D_{max}/D_{min}) \quad \text{Eq.1}$$

$$symmetry = 100 * |A_{left} - A_{right}|/(A_{left} + A_{right}) \quad \text{Eq. 2}$$

where, $D_{max}$ and $D_{min}$ are maximum and minimum readings on the profiles within 80% field width, and $A_{left}$ and $A_{right}$ are the areas under the curve within 50% field width split at the measured field center.

### *2.G. 2D imaging at the treatment position*

The kV-kV imaging system provides planar radiographic verification at the treatment position, analogous in geometry to robotic radiosurgery kV-kV tracking systems, to confirm that the patient has not moved between the upright-CT imaging position and the treatment position. Imager isocenter and localization accuracy were evaluated by aligning a cubic phantom with a radiopaque marker to the beam isocenter using upright CT, then acquiring simultaneous kV images from both panels and recording the offsets from the CT-defined isocenter. Auto-registration fidelity was assessed by applying a known 1.0 cm offset to the radiopaque marker phantom used for isocenter localization, in each of the lateral, longitudinal, and vertical directions independently, and comparing the software-recovered offset to the applied offset. Image quality was assessed with the kV-QA phantom (Sun Nuclear, a Mirion Medical Company, Melbourne, FL), evaluating spatial resolution in terms of modulation transfer function (MTF), contrast, noise, and uniformity. Imaging dose was measured for each clinical kVp/mAs combination using a RaySafe X2 calibrated dosimeter (Unfors RaySafe AB, Hovås, Sweden). Commissioning of the upright CT scanner itself, including CT dose index, image quality, and CT-number-to-stopping-power calibration, is reported separately[49] and is not included here.

### *2.H. Others*

The water-equivalent thickness (WET) of every material present in the clinical beam path was measured at 227 MeV using the IBA Zebra MLIC, comparing the R80 of the field delivered with and without the test object in the beam path. This characterization included the UPP backrest (on- and off-axis), the head-and-neck immobilization board (both from CQ Medical, Avondale, PA), an abdominal belt, a vacuum bag, an in-house pediatric chair backrest and headrest, and solid water slabs.

## 3. Results

### *3.A. Safety systems*

Safety-interlock testing confirmed that all systems (warning lights, emergency-stop functionality, motion-enable controls, door interlocks, last man out switches, touch sensors, patient audio-visual monitoring, radiation monitors, dose reporting and beam monitor ion chamber agreement) were functional.

### *3.B. Non-primary radiation*

Non-primary radiation, measured 100 cm laterally on either side of the phantom with a neutron bubble detector, a Fluke survey meter, and a Swendi detector, was below 0.01% of the delivered dose for all three detector types in both directions. The measured results with all three detectors are shown in Figure 2A.

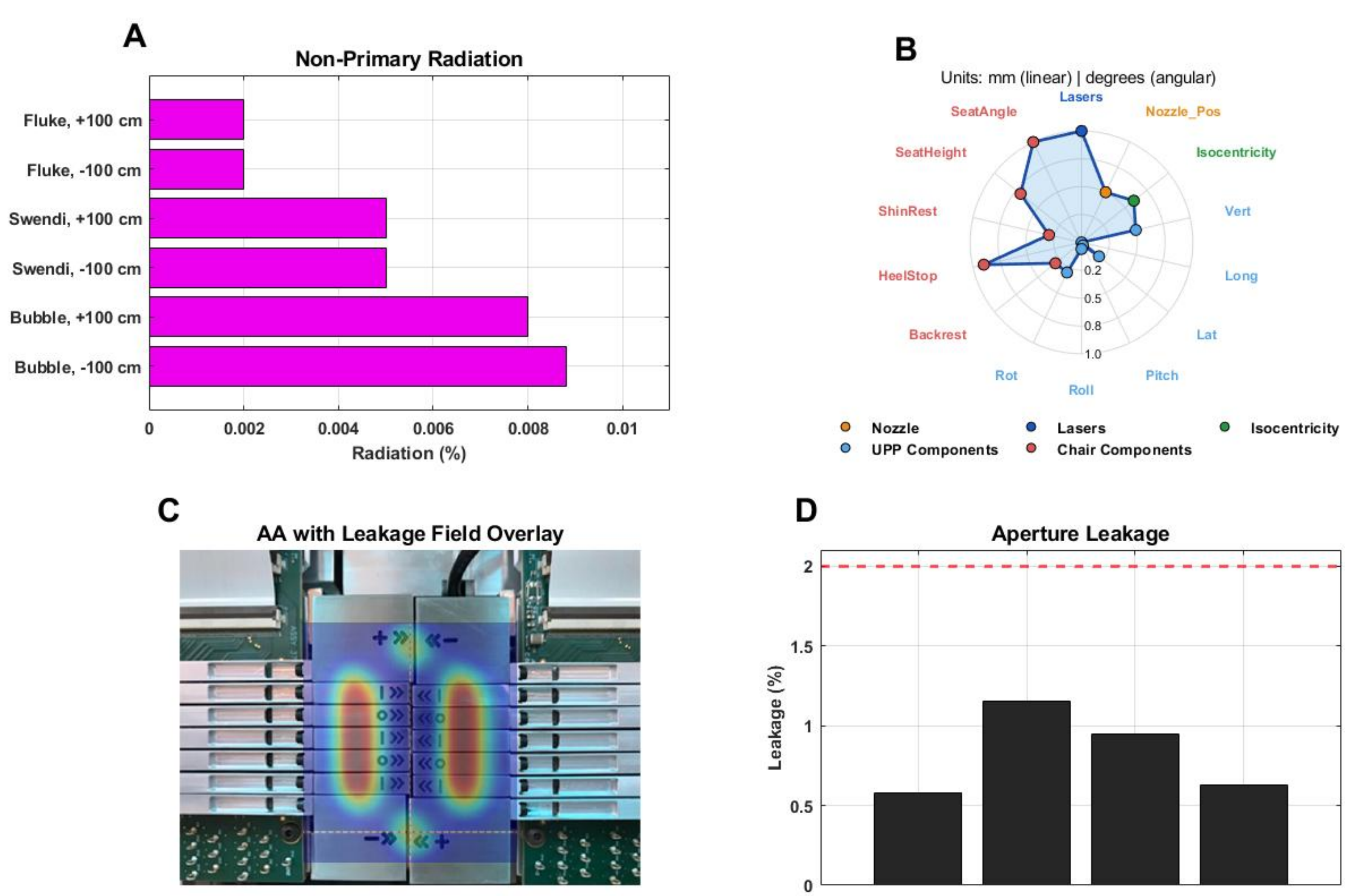

**FIGURE 2.** (A) Non-primary radiation at 100 cm laterally as a percentage of the delivered dose. (B) Spider plot of laser alignment, nozzle position, and isocentricity, together with UPP 6-DOF and chair-component accuracy. Values are in mm for linear measurements and degrees for angular measurements. (C) For illustration only: overlay of the AA assembly with the field used for the AA leakage test. (D) AA interleaf leakage at the jaw bevel and leaf tongue-and-groove interlock, relative to the 2% tolerance (dashed line).

### *3.C. Adaptive aperture*

AA leaf position accuracy was within ±1 mm of the nominal target position at all six tested positions. Dosimetric AA leaf positioning, evaluated with the automated picket-fence-style method, showed leaf edge position deviations from the strip mean of −0.2 to +0.2 mm in the X-narrow direction and −0.3 to +0.5 mm in the Y-narrow direction (Figure 3A, B). Interleaf leakage, measured separately at the jaw interlocking bevel and the leaf tongue-and-groove interlock at maximum energy, ranged from approximately 0.6% (jaws) to approximately 1.2% (leaves) of the uncollimated maximum dose (Figure 2D), below the ≤2% vendor specification and consistent with the <1.5% leakage previously reported for the same AA design on the gantry-mounted platform[10]. AA penumbra reduction was confirmed across all tested energy, field-size, and air-gap combinations, with the largest absolute reduction observed at the lowest energy (49.3 MeV) and the smallest air gap (8.6 cm; up to 14.6 mm), and progressively smaller reductions at higher energy and larger air gap (as low as 0.05-0.28 mm at 227.1 MeV, 38.6 cm) (Figure 3C). This trend, greater sharpening at smaller air gaps is consistent with prior reports for this AA design[10].

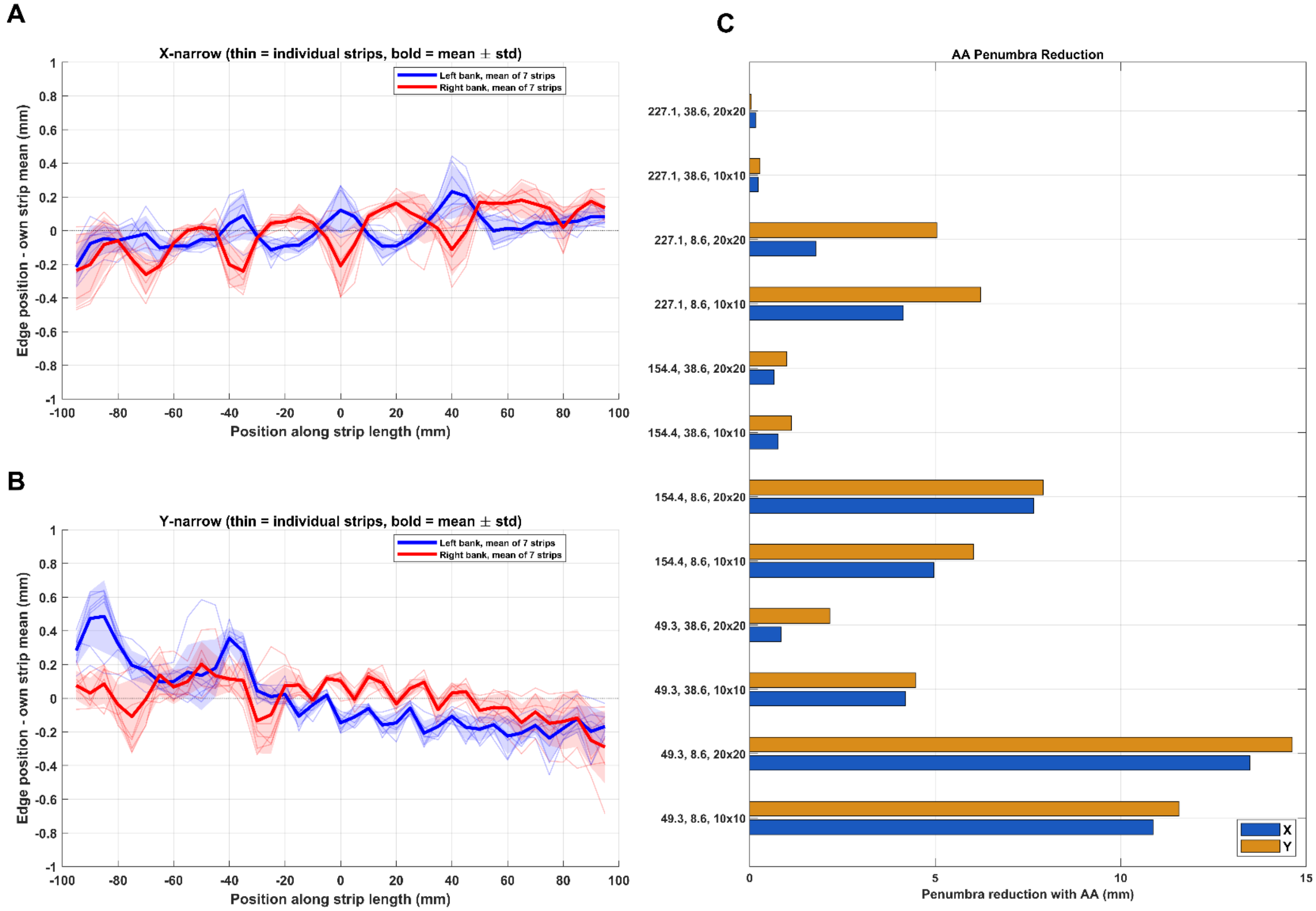

**FIGURE 3.** AA performance. (A, B) Dosimetric leaf-position verification, X and Y orientation: edge position along the 200 mm strip length, all 7 strips overlaid per bank. (C) Penumbra reduction with AA across energy, air gap, and field size. AA leaf positioning was accurate and reproducible along the full strip length, and the AA reduced penumbra by up to 14.6 mm, with the greatest sharpening at the smallest air gap.

### *3.D. Mechanical commissioning*

All UPP, chair-component, and nozzle positions agreed with control-system indicators within their respective tolerances across the full clinical range of each component, with chair components generally showing larger, though still within-tolerance, deviations than the UPP components or nozzle position (Figure 2B). The 6-DOF UPP translational accuracy was within ±1 mm in the lateral, longitudinal, vertical directions, and rotational, pitch, and roll accuracy were within ±0.3°. All seven alignment lasers agreed with the imaging isocenter and with known UPP translational offsets to within ±1 mm (Figure 2B). Isocentricity was within 0.6 mm at all tested rotation angles, shown in Figure 2B.

### *3.E. Beam commissioning and validation*

Across the 12 commissioning energies (49.3-227.1 MeV), R90 ranged from 21.97 to 321.75 mm and R80 from 22.88 to 322.71 mm for water tank measurements, agreeing with the IBA Zebra MLIC (R90 21.5-321.8 mm, R80 22.5-322.8 mm) to within 1 mm at all 12 energies. The 80%-80% peak width ranged from 8.32 to 8.83 mm, and the 20%-80% distal falloff ranged from 4.50 to 4.66 mm, shown in Figure 4A.

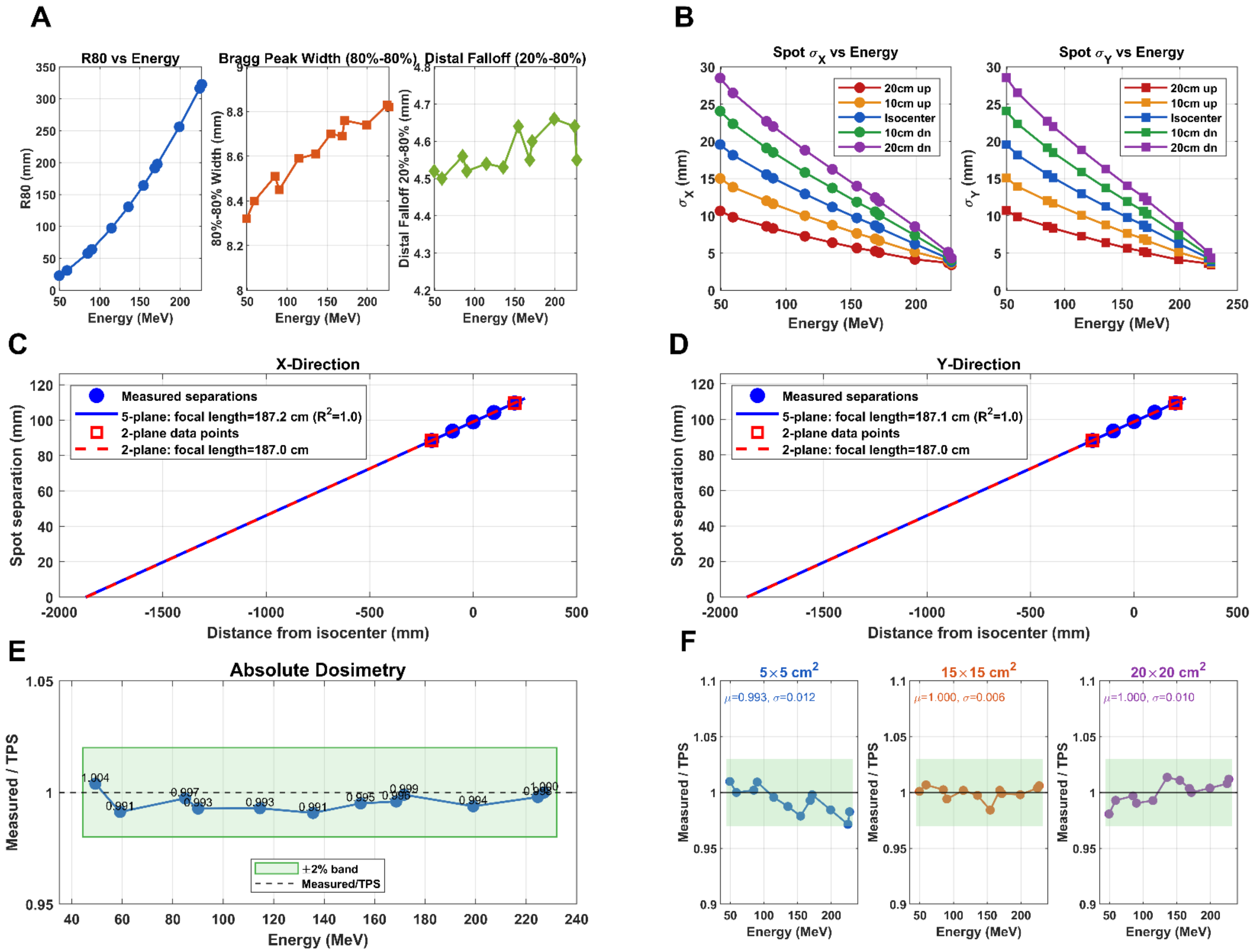

**FIGURE 4.** (A) IDD parameters: R80, 80%-80% peak width, and 20%-80% distal falloff vs. energy. (B) Spot $\sigma_X$ and $\sigma_Y$ vs. energy at five positions along the beam axis (20 and 10 cm upstream of isocenter, isocenter, 10 and 20 cm downstream). (C) X-direction focal length: spot separation vs. distance from isocenter, with 5-plane and 2-plane linear extrapolation to zero separation. (D) Y-direction focal length, as in (C). (E) Absolute dosimetry: measured/TPS dose ratio vs. energy at the reference 10×10 cm$^2$ field size. (F) Output for 5×5, 15×15, and 20×20 cm$^2$ fields.

Spot sigma at isocenter ranged from 3.83 mm (227.1 MeV) to 19.58 mm (49.3 MeV) and increased with distance downstream from isocenter at every energy, 3.4-10.7 mm at 20 cm upstream vs. 4.3-28.5 mm at 20 cm downstream across the full energy range, as shown in Figure 4B.

The five-plane linear fit gave a focal length of 187.2 cm ($R^2$ = 1.0) in the X-direction (Figure 4C) and 187.1 cm ($R^2$ = 1.0) in the Y-direction (Figure 4D). The two-plane cross-check gave concordant results (187.0 cm and 187.0 cm, respectively), agreeing with the five-plane fit to within 0.1% and 0.05%. Thus, the average focal length of 187 cm for both directions was adopted for beam modeling.

Absolute dose at the reference 10×10 cm$^2$ field size agreed with the TPS to within 1% at all 12 commissioning energies, Figure 4E. Output agreed with the TPS to within 3% across the three additional field sizes tested (5×5, 15×15, 20×20 cm$^2$) and all 12 energies, Figure 4F.

For the 25-spot position accuracy pattern, measured spot centroids deviated by up to 0.20 mm in X and 0.37 mm in Y (mean 0.07 mm and 0.13 mm, respectively), with corresponding spot sigmas of 4.06-4.30 mm (X) and 3.72-4.04 mm (Y) across the 20×20 cm$^2$ field at the unmodulated highest energy (Figure 5A). Monolayer field size, penumbra, flatness, and symmetry for the three representative energies at the smallest and largest air gap is shown in Figure 5B. SOBP field size agreed with the TPS to within 0.78 mm and lateral penumbra to within 0.98 mm for the target volumes and depths evaluated, Figure 5C. MU linearity at 227 MeV was within 1.2% across the full range tested (0.271-20 MU/spot; Figure 5D). The maximum achievable field size of 20×20 cm$^2$ was confirmed by the 50% isodose of a uniform scanned field.

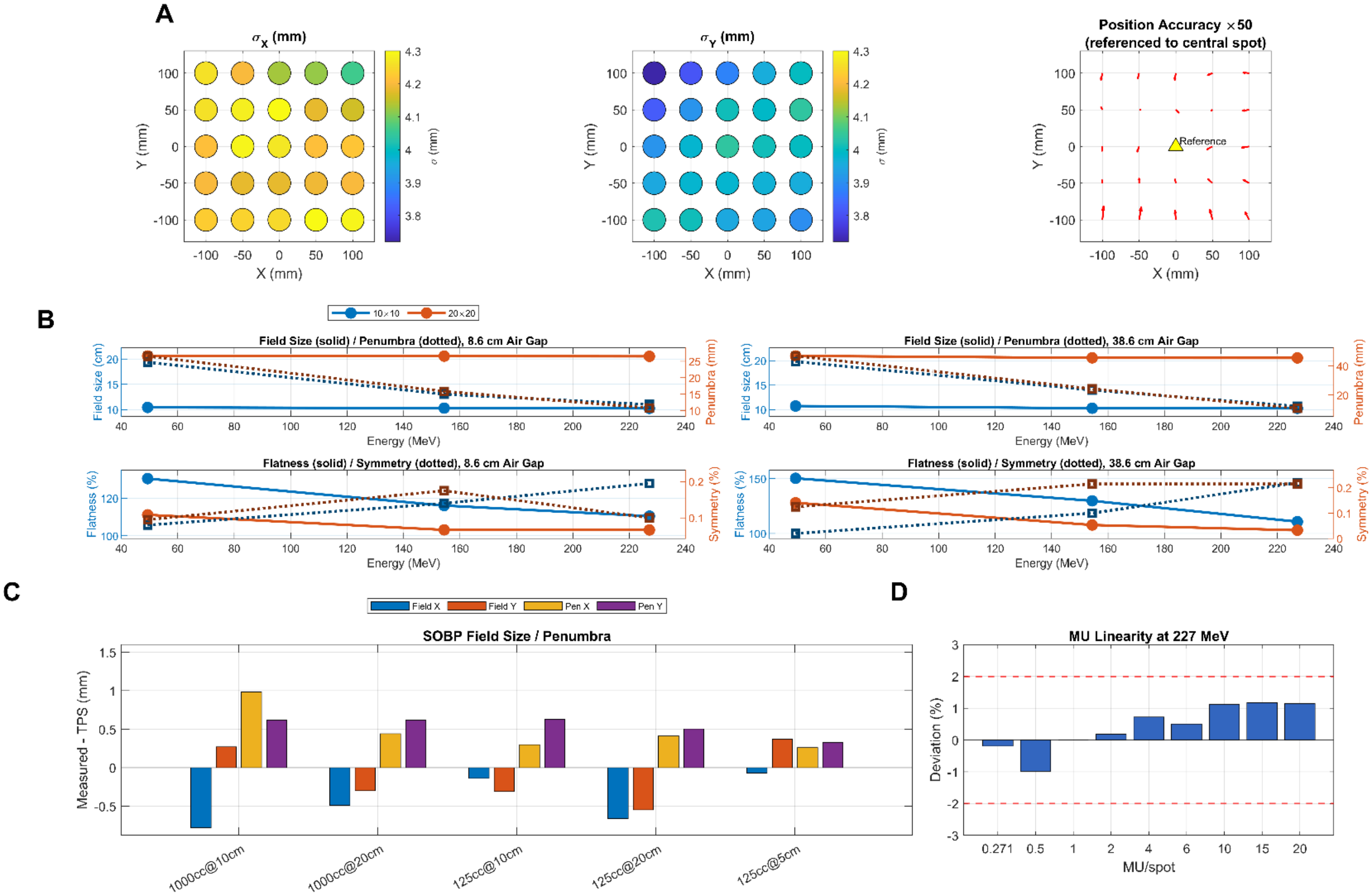

**FIGURE 5.** (A) Off-axis spot size ($\sigma_X$, $\sigma_Y$) across the 20×20 cm$^2$ field and spot position accuracy (arrows, ×50 exaggeration), referenced to the central spot; (B) monolayer field size, penumbra, flatness, and symmetry; (C) SOBP field size/penumbra (measured − TPS) across target volumes and depths; (D) MU linearity at 227 MeV.

Gamma passing rates for SOBP absolute-dose validation (open, static AA, and dynamic AA configurations) exceeded 91% at the 3%/2 mm criterion for all target volumes up to 1125 cm$^3$, and central-axis point dose agreed with the TPS to within 3% for every plan tested, Figure 6A. Site-specific validation plans for all seven anatomical sites (brain, C-shape, head-and-neck, pelvis, breast, prostate, and craniospinal irradiation) met the institutional 3%/2 mm gamma criterion (≥94%) and point-dose acceptance criteria at every measured depth and field, Figure 6B.

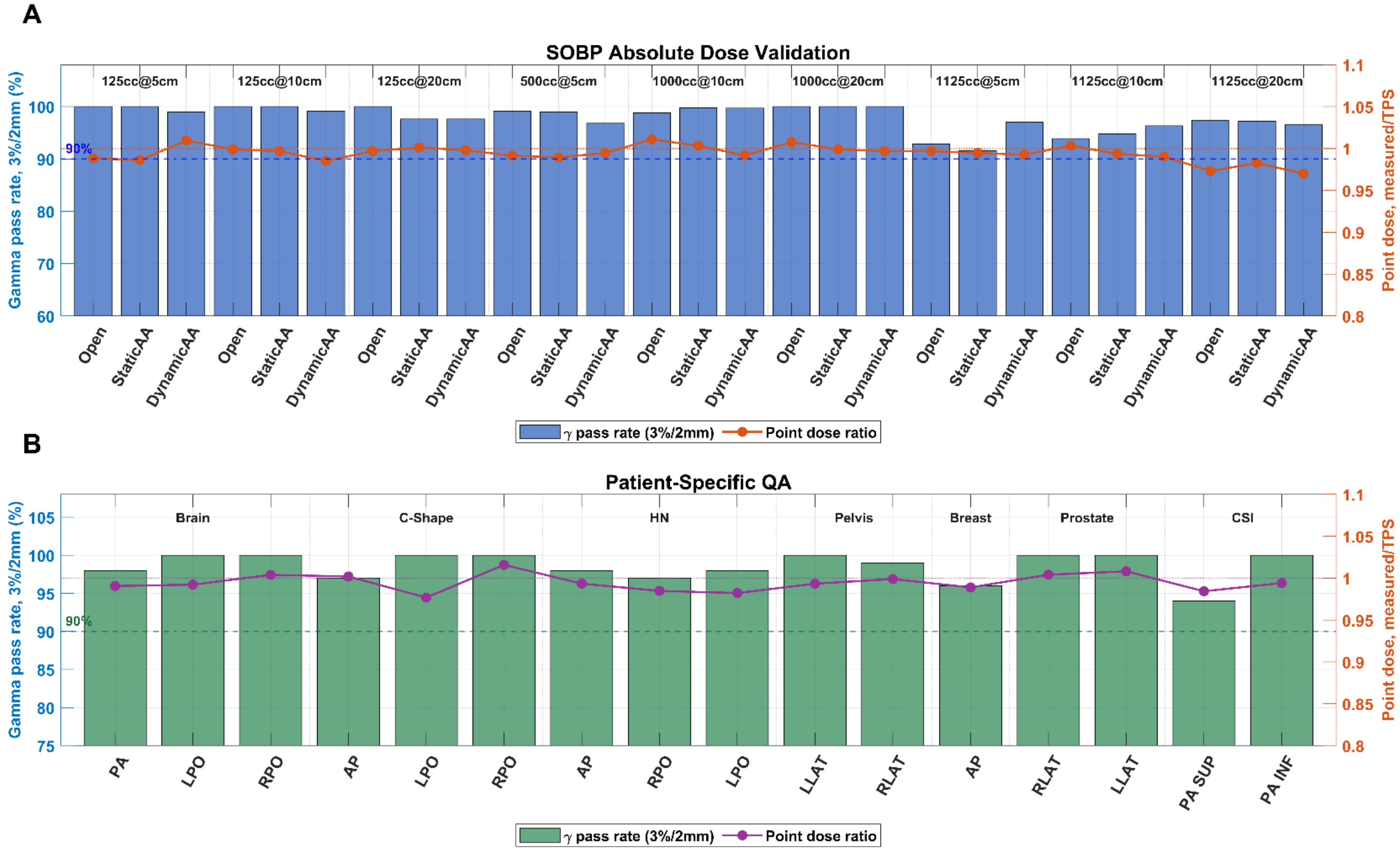

**FIGURE 6.** (A) Gamma pass rate (3%/2mm, bars, left axis) and central-axis point dose ratio (measured/TPS, markers, right axis) for SOBP absolute-dose validation plans (top; grouped by target volume and depth, labeled above each group, with open, static AA, and dynamic AA conditions shown per group) and (B) for site-specific validation plans (bottom; grouped by anatomical site, labeled above each group, with individual beam field IDs shown per group). Gamma passing rates exceeded 91% and point doses agreed with the TPS to within 3% across all SOBP and site-specific validation plans.

End-to-end testing on the in-house phantom demonstrated agreement between delivered and planned dose within 3% for all beams and for the composite plan (Figure 7A), and axial film gamma passing rates of 95% and 91%, both exceeding the 90% institutional criterion (Figure 7B). The independent IROC audit confirmed acceptable machine output, within 2% (ratio of absorbed dose determined by IROC to that stated by us was 1.02). For the anthropomorphic head-and-neck phantom irradiation (approximately 6.6 Gy(RBE) prescribed to the primary planning target volume), TLD point-dose ratios (measured/reported) ranged from 0.96 to 1.09 across the target and three organs-at-risk (Figure 7C), and GafChromic film gamma passing rates were 92% (axial plane) and 94% (sagittal plane), both exceeding the IROC acceptance threshold of 85% (Figure 7D).

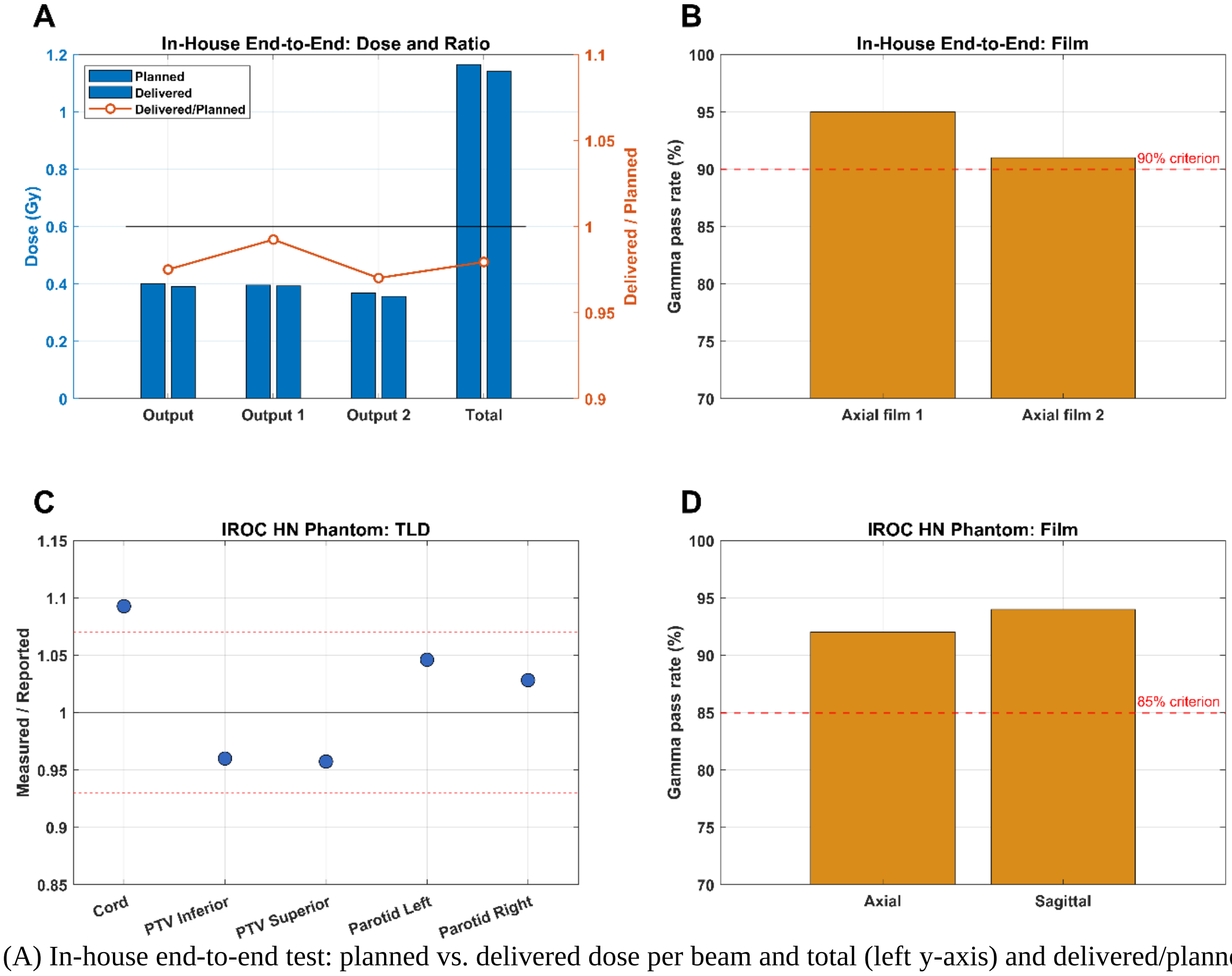

**FIGURE 7.** (A) In-house end-to-end test: planned vs. delivered dose per beam and total (left y-axis) and delivered/planned ratio (right y-axis). (B) In-house end-to-end test: axial film gamma pass rate against a 90% criterion. (C) IROC anthropomorphic head-and-neck phantom: TLD point-dose ratios (measured/reported) for the PTV and organs at risk (dotted lines denote ±7%). (D) IROC anthropomorphic head-and-neck phantom: film gamma passing rates (axial, sagittal) against the 85% IROC acceptance criterion.

### *3.F. 2D imaging at the treatment position*

The kV-kV imaging isocenter agreed with the upright-CT-defined isocenter to within 1 mm in all three directions (lateral -0.7 mm, longitudinal -0.3 mm, vertical 0.5 mm). Auto-registration fidelity was verified by the software-recovered offsets of 1.02 cm (lateral), 1.00 cm (longitudinal), and 0.94 cm (vertical), Table 2. Imaging dose ranged from 0.074 mGy at 80 kVp to 0.198 mGy at 120 kVp (200 mA, 1 s exposure). Image quality was consistent between the two imaging panels, with MTF50 of 1.59-1.63 lp/mm, uniformity of 99.7%-99.8%, and contrast of 0.70-0.74. Noise was slightly higher for imager 2 compared to imager 1 (35.1 vs. 25.5 HU), Table 2.

**TABLE 2.** kV-kV imaging system: imaging dose, image quality, and geometric accuracy.

| **Imaging dose** | | | |
|---|---|---|---|
| **Nominal kVp** | **mA** | **Exposure (s)** | **Dose (mGy)** |
| 80 | 200 | 1 | 0.074 |
| 90 | 200 | 1 | 0.116 |
| 100 | 200 | 1 | 0.142 |
| 110 | 200 | 1 | 0.169 |
| 120 | 200 | 1 | 0.198 |
| **Image quality** | | | |
| **Metric** | **Imager 1** | **Imager 2** | |
| MTF50 (lp/mm) | 1.59 | 1.63 | |
| MTF40 (lp/mm) | 1.86 | 1.91 | |

| MTF30 (lp/mm) | 2.15 | 2.20 | |
|---|---|---|---|
| Uniformity (%) | 99.76 | 99.69 | |
| Contrast | 0.70 | 0.74 | |
| Noise (HU) | 25.5 | 35.1 | |
| **Isocentricity offset (kV-kV vs. upright-CT-defined isocenter)** | | | |
| **Direction** | **Offset (mm)** | | |
| Lateral | -0.7 | | |
| Longitudinal | -0.3 | | |
| Vertical | 0.5 | | |
| **Auto-registration fidelity** | | | |
| **Applied offset** | **Direction** | **Recovered offset** | |
| 1.0 cm | Lateral | 1.02 cm | |
| 1.0 cm | Longitudinal | 1.00 cm | |
| 1.0 cm | Vertical | 0.94 cm | |

### *3.G. Others*

The measured WET of clinical beam-path hardware ranged from 2.1 mm (pediatric chair headrest) to 10.45 mm (average for 1-cm-thick solid water slabs used for beam-path verification). The UPP backrest from CQ Medical measured 4.6-4.7 mm both on-axis and ±10 cm off-axis, the CQ Medical head-and-neck immobilization backboard 4.5 mm, the abdominal belt 3.8 mm, and the vacuum bag 3.4 mm; the pediatric chair backrest and headrest each measured 2.1-2.2 mm at the tested positions.

## 4. Discussion

The commissioning results presented here demonstrate that Stanford PA-1 satisfies all performance criteria across mechanical, dosimetric, and TPS domains, supporting the clinical release of the S250-FIT platform. Several findings merit specific discussion within the context of the system's novel fixed-beam, gantry-free architecture and its boron-carbide-based ES.

### *4.A. Practical challenges of upright beam commissioning*

A consequence of the fixed-beam architecture is that every detector and phantom had to be reoriented to align with the horizontal beam, equivalent to gantry 90° on a rotating-gantry system. For some detectors, this is a simple substitution: mounting vertically rather than horizontally. For others, the modification was more complex, requiring custom-designed adapters.

IDD and output measurements presented several practical challenges specific to the fixed-beam geometry. Because the beam enters through a side window of the water tank, accurate knowledge of the WET of the side window, the reference detector, and any water gap between them becomes critical to correctly referencing the measured curve to true zero depth. We characterized these WETs with a calibrated MLIC and used vendor-provided precision calibration gauges to measure the physical water gap between the side window and the detector surface. A second, largely mechanical challenge arose from the 3D water tank's positioning platform, which is normally located beneath the tank. With the UPP occupying that space in the upright geometry, the tank instead required a custom-designed support platform (see Supplementary Document), and the tank's water-reservoir hoses had to be lengthened accordingly. A filled 3D tank is also substantially heavier once mounted on the UPP's 6-DOF motion system, a consideration that should be factored into the mechanical design and motion-safety margins for any similarly configured system. For practical day-to-day commissioning, we found a 1D water tank more tractable than the 3D system, at the cost of two limitations: field size is constrained to approximately 10×10 $cm^2$, adequate for absolute output verification but not for field-size-specific output measurements, and the 1D geometry does not support lateral profile scanning.

Spot profile measurements with the IBA Lynx required a custom flat platform interlocking with the UPP to hold the detector in the beam path. Our first iteration of the platform was rectangular, and its diagonal extent exceeded the bore diameter of the upright CT, creating a collision risk that prevented CT-based alignment verification of the detector in that configuration. We are working on the second iteration, redesigned to remain within the bore envelope. Centers can also use a Phoenix detector instead on the same custom platform described above for the 3D water tank, with minor modifications.

Isocentricity measurements used an XRV scintillator, which is normally positioned flat on a couch in a supine geometry. A custom vertical adapter that mounts on the backboard, developed together with the manufacturer (Logos Systems International), was required to hold the detector upright in the fixed horizontal beam. A further practical consideration is that, depending on the UPP rotation angles, the beam entrance may be partially obstructed by the backboard itself, broadening the measured spot on that side. While this is a minor effect, we recommend mounting the adapter as high as practical on the backboard so that the detector cone extends clear of the backboard.

Beyond detector positioning, the nozzle exit window introduces a further modeling consideration. Of the three measurement types used to build the beam model, the IDD and spot profile were acquired with the window removed, while the absolute output used for dose calibration was acquired with the window in place. Commissioning teams should track this distinction carefully, since the window's water-equivalent thickness must be accounted for exactly once; inconsistent treatment risks double-counting or omitting it entirely. All subsequent verification measurements in this work, consistent with actual patient treatment, were performed with the window in place.

Measurement of imaging dose from the kV-kV system with a RaySafe X2 probe presented a related geometric challenge because the kV X-ray sources are installed obliquely (see Supplementary Document) relative to the UPP. Correctly positioning the probe perpendicular to each source may require coordinated use of all 6-DOFs of the UPP. Centers commissioning similar oblique kV-kV geometries should anticipate this complexity for image quality measurements as well. As an alternative, the image-quality phantom can instead be positioned directly on the imager panel itself, with the baselines characterized at that setup.

2D planar dose measurements with the Octavius 1500XDR array were calibrated using an SOBP field rather than a single energy layer, matching the SOBP-based commissioning condition used by the IROC for their TLDs. We also observed that Octavius readouts became increasingly noisy for longer acquisitions, particularly for large fields delivered with many energy layers and spots. Re-zeroing the detector between measurements mitigated this to some degree. This noise was most pronounced for fields delivered with the AA (static or dynamic), which require substantially longer delivery times than open fields. We attribute this, in part, to stray radiation generated as protons strike the AA leaves, including a neutron component, though we did not isolate the mechanism directly in this work.

To avoid the need for physical rulers or tape measures at each tested position, we found laser distance measurers a practical and efficient tool, especially for CT height verification.

Photographs of each setup described above are provided in the Supplementary Document, both as a record of the geometry used for each measurement and as a reference for other centers commissioning similar systems.

### *4.B. Mevion S250-FIT-specific considerations*

First, a direct comparison of spot sigma at isocenter with the previously reported S250i platform, Figure 8, shows that the S250-FIT consistently produces smaller spots across the shared energy range, with the difference increasing up to 14% as energy decreases. It is likely that boron carbide achieves the same water-equivalent range modulation with a thinner physical plate stack than the all-Lexan S250i ES. A similar finding was noted by Zhao et al. in their simulation study[58].

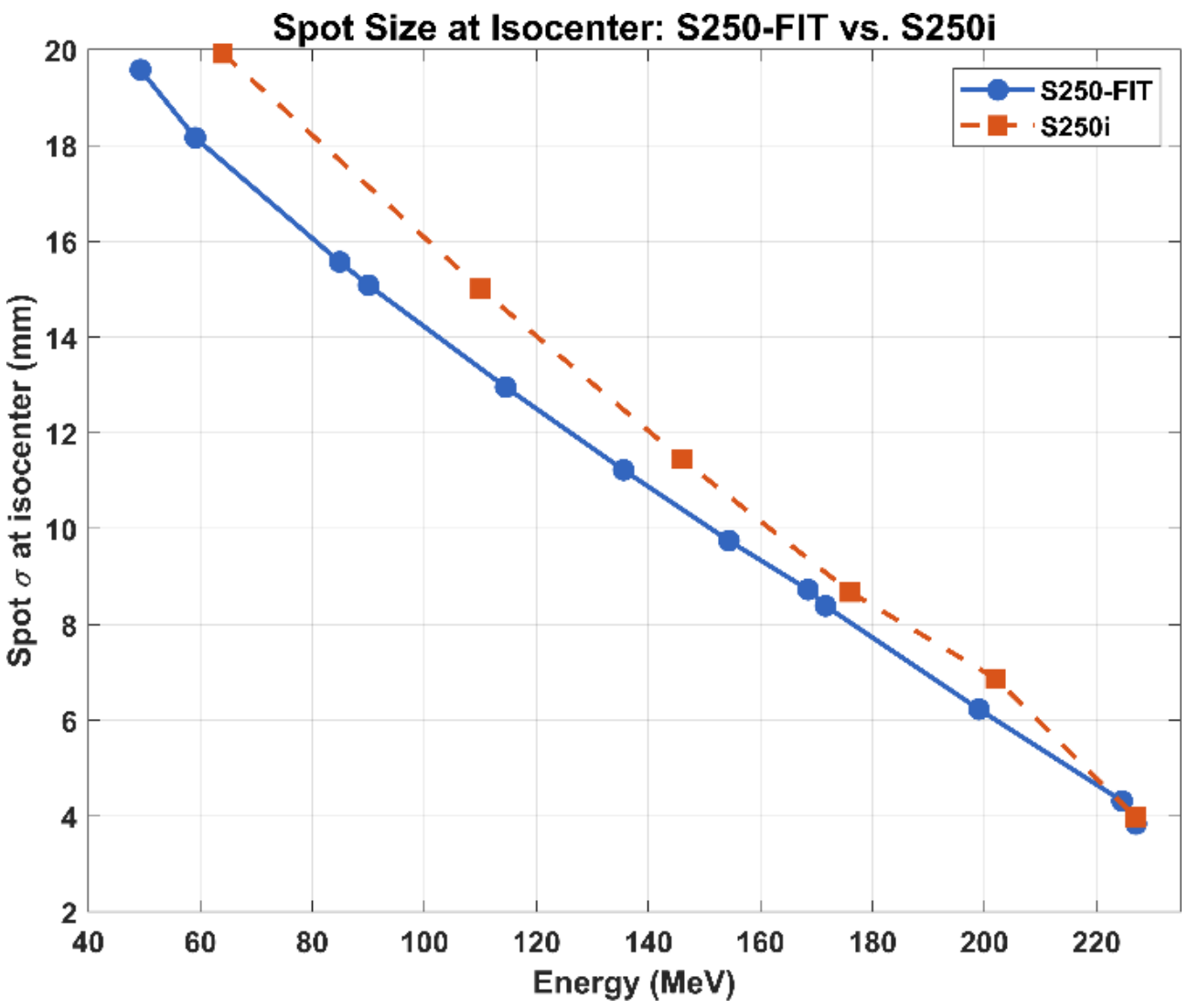

**FIGURE 8.** Spot σ at isocenter as a function of energy for the S250-FIT and the previously reported S250i[10] platform. The S250-FIT produced smaller spot sizes than the S250i across the shared energy range, consistent with reduced scattering from the thinner boron carbide energy selector.

Second, both the 80%-80% Bragg peak width (8.3-8.8 mm) and the 20%-80% distal falloff (4.5-4.7 mm) vary only modestly across the clinical energy range we tested, in contrast to the behavior reported[47] for beamline-based systems that refocus the beam with quadrupole magnets after a separate energy-selection system, where the Bragg peak narrows progressively at lower energy. This mirrors the trend previously reported for the gantry-mounted S250i platform[10,11], which uses the same class of ES-based range modulation without downstream beam refocusing.

Third, it is worth noting that the chair axes (seat height, seat pan angle, shin rest, heel stop, and backrest angle) serve a fundamentally different clinical role than the 6-DOF platform axes. The chair axes establish and fix the patient's immobilized position for a given simulation setup. Once set, they remain unchanged for the duration of a course of treatment, so their reproducibility matters clinically. The 6-DOF platform axes, in contrast, are used daily to reposition the patient relative to the fixed beam following image-based registration, and their movement accuracy directly determines positioning accuracy at each fraction. Both properties were evaluated in this work and met their respective tolerances, but they warrant distinct consideration in future QA program design.

### *4.C. Limitations*

The agreement between institutional measurements and the independent IROC audit, both for absolute machine output and for the anthropomorphic head-and-neck end-to-end irradiation, provides external validation of the commissioning results reported here. The one exception was the spinal cord TLD, which exceeded the ±7% IROC acceptance range (measured/reported=1.09). Given the steep dose gradient at the cord in this head-and-neck plan, a small physical offset in TLD positioning can translate into a comparatively large dose difference.

Consistent with the scope defined at the outset, facility shielding design and survey, and commissioning of the in-room upright CT scanner are reported separately and are not included here. The present report addresses only the supplementary kV-kV system used for treatment-position verification. A related limitation of the current clinical workflow is that volumetric CT imaging cannot be acquired once the patient has been moved to the treatment position. The kV-kV system therefore serves as the sole means of confirming that no patient motion occurred between CT-based alignment and treatment. A further limitation is that the current system does not support shift-and-treat based on the kV-kV images themselves: if a kV-kV pair reveals a discrepancy, the only available corrective action is to return the patient to the imaging position for re-alignment and re-acquisition of CT. While this constrains workflow efficiency, it may also carry an incidental benefit, since 2D planar imaging provides comparatively limited positional information relative to volumetric CT, and shifting treatment based on 2D imaging alone after 3D alignment has already been performed could introduce risk.

As with any first-in-kind clinical system, longer-term follow-up of mechanical and dosimetric stability, and clinical outcomes data from the first treated patients, will be important complements to the commissioning reported here.

## 5. Conclusions

We have reported the commissioning of Stanford Proton Accelerator 1 (PA-1), a Mevion S250-FIT and the world's first clinical implementation of an ultra-compact, gantry-free, fixed-beam upright PBS system based on a superconducting synchrocyclotron. Comprehensive mechanical, dosimetric, and TPS commissioning, together with an independent external audit, demonstrate that the system meets institutional criteria for clinical use. The first pediatric[59] and adult patients were treated on this system on June 4, 2026. Because this fixed-beam, UPP-based architecture lacks a direct precedent among compact synchrocyclotron PBS systems, it introduces unique considerations for commissioning methodology and beam characterization. With additional S250-FIT units reportedly in installation or planning worldwide, the methodology detailed herein is offered as a reference for future commissioning of upright, gantry-free proton systems.

## Acknowledgments

The authors thank the Mevion Medical Systems and Leo Cancer Care engineering and physics teams for their support during acceptance and commissioning.

**Conflict of Interest Statement:** BWL has received lecture honoraria from Mevion Medical Systems. MRP and SG are employees of Mevion Medical Systems and Leo Cancer Care, respectively. The other authors have no conflicts to disclose.